\def\hcorrection#1{\advance\hoffset by #1 }
\def\vcorrection#1{\advance\voffset by #1 }

\documentstyle[11pt]{article}

\vcorrection{-0.70in}
\hcorrection{-0.40in}
\begin{document}

\title{Analysis and boundary condition 
of the lattice Boltzmann BGK model with two velocity components}
\author{Xiaoyi He \thanks{%
Center for Nonlinear Studies, Los Alamos National Lab, Los Alamos,
NM 87545} \thanks{%
Theoretical Biology and Biophysics Group, Los Alamos National Lab} 
and Qisu Zou \thanks{%
 Theoretical Division, Los Alamos National Lab} 
\thanks{%
 Dept. of Math., Kansas State University, Manhattan, KS 66506} 
} 
\date{}
\maketitle

\begin{abstract}
In this paper, we study the two dimensional 
lattice Boltzmann 
BGK model (LBGK) by analytically solving a simple flow in a 2~-D channel.
The flow is driven by the movement of upper boundary  
with vertical injection fluid at the porous boundaries. 
The velocity profile is shown
to satisfy a  second-order finite-difference  form of the simplified
incompressible Navier-Stokes equation. With the analysis, different
 boundary conditions can be studied  theoretically. A momentum exchange
principle is also revealed at the boundaries. A general 
boundary condition for
any given velocity boundary is proposed based on  the analysis. 
\end{abstract}

\newpage

\section{Introduction}

The lattice Boltzmann equation (LBE) method has become a promising tool for
simulation of transport phenomena in recent years. Great success has been
achieved in applying LBE for various flow problems such as
 hydrodynamics
\cite{sbh,cws,ma,mmcm}, flow
through porous media \cite{gens1,gens2,daryl}, magnetohydrodynamics 
\cite{ccma,svb,mcm},
multiphase flow \cite{gens1,gens2,gce,shan1,shan2,roth3}, 
reaction-diffusion equation \cite{dcd}, and particle suspensions
\cite{ladd}. 
Compared to its precursor, the lattice gas automata (LGA), LBE method
is more computationally efficient using current parallel computers, and
some artifacts like non-Galilean invariance in LGA can be eliminated in LBE. 
Careful qualitative  comparisons of LBE with traditional computational 
fluid dynamics methods showed  that the method  is accurate, and capable of
simulating complex phenomena \cite{ma,mmcm,marc,hou1,hou2}. 

Although it has been proved \cite{hchen,qian2,qian3,hou1} that the LBE 
recovers the Navier-Stokes 
equation with a second-order of accuracy in space and time
in the interior of flow domain, the real hydrodynamic boundary
conditions have not been well studied. The no-slip condition on the
wall is one of the examples. The classical model for the non-moving wall 
boundary
condition in LBE is the so-called bounce-back rule borrowed from the LGA
method. Under the bounce-back rule, all the particles colliding with walls
bounce back to the flow domain in the same direction, 
Theoretical discussion and computational experience indicate that
the bounce-back rule actually gives a zero velocity half way between the
bounce back row and the first row in the flow and it
introduces an error of first-order in lattice spacing for LGA and LBE
\cite{corn,ginz,hou1}. 
To solve this problem, various boundary conditions \cite{nobel,bob}
 have been proposed to replace  the
bounce-back rule and progress has been made.
In \cite{nobel}, a boundary condition for the triangular (FHP) LBGK model
was proposed for any given velocity boundary, the boundary 
condition generated  results of machine accuracy for plane Poiseuille
flow. 
In \cite{bob}, a non-slip boundary condition and  prescribed pressure
or velocity inlet condition 
for the 3-D 15-velocity direction LBGK
model were proposed and results of good accuracy for various flows 
are achieved. 
However, due to the lack of the fundamental physical reasoning and  the
lack of mathematical analysis, these schemes
have not revealed  the general nature of boundary conditions.

Recently we have developed  \cite{he1}
a new technique to analytically solve the 
LBGK equation for 2-D Poiseuille flow and Couette flow.  This technique
provides us a useful tool to analyze
the error generated by various boundary conditions.  In this
study, we will extend our effort to include a flow with
velocities in both directions. In addition, a general boundary  condition
for any given velocity straight boundary is proposed.

\section{Governing Equation}
In this study, we will only use the square lattice LBGK model.  The 
procedure can be easily extended to the triangular lattice model, 
for which study
is easier due to the smaller number of velocity directions than that of
the square model.

The model is expressed as:
\begin{equation}
f_i({\bf x}+\delta {\bf e}_i, t+\delta)-f_i({\bf x},t)=
 -\frac 1{\tau}[f_i({\bf x},t)-f_i^{(eq)}({\bf x},t)], \ \ i=0,1,...,8,
\label{eq:lbgk}
\end{equation}
where  the equation is written in physical units.
Both the time step and the lattice spacing have the value of $\delta$
in physical units.
$f_{i}({\bf x},t) $ is the density distribution
function along the direction ${\bf e}_i$ at (${\bf x}, t$).  The 
particle speed ${\bf e}_i$ are given by
 ${\bf e}_{i} = (\cos(\pi(i-1)/2), \sin(\pi(i-1)/2), i = 1,2,3,4$, and
${\bf e}_{i} = \sqrt{2} (\cos(\pi(i-4-\frac{1}{2})/2),
\sin(\pi(i-4-\frac{1}{2})/2), i = 5,6,7,8$.
Rest particles of type 0 with ${\bf e}_{0} = 0$ is also allowed.
The  right hand side represents the collision term
 and $\tau $ is the single relaxation time which controls the rate
of approach to equilibrium.
The density per node, $ \rho $, and the macroscopic flow velocity,
 ${\bf u} = (u, v)$,
are defined in terms of the particle distribution function by
\begin{equation}
 \sum_{i=0}^8 f_i=\rho,  \ \ \ \
 \sum_{i=1}^8 f_i {\bf e}_i =\rho {\bf u}.
\label{eq:dens}
\end{equation}
The equilibrium distribution functions 
$f_i^{(eq)}({\bf x},t) $ 
depend only on local density and
velocity and they
can be chosen in the following form (the model d2q9 \cite{qian2}):
\begin{eqnarray}
f_{0}^{(eq)}&=&\frac{4}{9}\rho[1-\frac{3}{2}{\bf u}\cdot{\bf u}],
\mbox{\hspace{1.65 in}}
\nonumber\\
f_{i}^{(eq)}&=&\frac{1}{9}\rho[1+3({\bf e}_{i}\cdot
{\bf u})+\frac{9}{2}({\bf e}_{i}\cdot {\bf u})^2-\frac{3}{2}{\bf u}\cdot
{\bf u}],\;\; i =  1,2,3,4 \\
f_{i}^{(eq)}&=&\frac{1}{36}\rho[1+3({\bf e}_{i}\cdot
{\bf u})+\frac{9}{2}({\bf e}_{i}\cdot {\bf u})^2-\frac{3}{2}{\bf u}\cdot
{\bf u}], \;\; i =  5,6,7,8 . \nonumber
\label{eq:equil}
\end{eqnarray}

Assume the flow is steady and 
\begin{equation}
\frac{\partial u}{\partial x}=0, \  \ \frac{\partial v}{\partial x}=0,
 \  \ \rho = \mbox{const} ,  
\label{eq:stead}
\end{equation}
then $f_i({\bf x},t)$ is only a function of $y$. This happens when
the flow is driven by boundaries moving in the $x$-direction with
injection in the $y$-direction  from the porous boundaries (see Fig.~1).
From Eq.~(\ref{eq:lbgk}) we have  
\begin{eqnarray}
f_0^j&=&\frac{4\rho }9[1-1.5(u_j^2+v_j^2)] \nonumber \\ 
f_1^j&=&\frac \rho 9[1+3u_j+3u_j^2-1.5v_j^2] \nonumber \\
f_2^j&=&\frac \rho {9\tau }[1+3v_{j-1}+3v_{j-1}^2-1.5u_{j-1}^2]
+(1-\frac 1\tau )\ f_2^{j-1} \nonumber \\ 
f_3^j&=&\frac \rho 9[1-3u_j+3u_j^2-1.5v_j^2] \nonumber \\ 
f_4^j&=&\frac \rho {9\tau }[1-3v_{j+1}+3v_{j+1}^2-1.5u_{j+1}^2]
+(1-\frac 1\tau )\ f_4^{j+1} \\ 
f_5^j&=&\frac \rho {36\tau }[1+3u_{j-1}+3v_{j-1}+3u_{j-1}^2+3v_{j-1}^2+9u_{j-1}v_{j-1}]
+(1-\frac 1\tau )\ f_5^{j-1} \nonumber \\
f_6^j&=&\frac \rho {36\tau }[1-3u_{j-1}+3v_{j-1}+3u_{j-1}^2+3v_{j-1}^2-9u_{j-1}v_{j-1}]
+(1-\frac 1\tau )\ f_6^{j-1} \nonumber \\
f_7^j&=&\frac \rho {36\tau }[1-3u_{j+1}-3v_{j+1}+3u_{j+1}^2+3v_{j+1}^2+9u_{j+1}v_{j+1}]
+(1-\frac 1\tau )\ f_7^{j+1} \nonumber \\
f_8^j&=&\frac \rho {36\tau }[1+3u_{j+1}-3v_{j+1}+3u_{j+1}^2+3v_{j+1}^2-9u_{j+1}v_{j+1}]
+(1-\frac 1\tau )\ f_8^{j+1}, \nonumber
\label{eq:evo}
\end{eqnarray}
where $ f_i^j $ stands for the density distribution along the direction 
$ {\bf e}_i $ at $ y=j \delta $.

According to Eqs.~(\ref{eq:dens}, \ref{eq:evo}), the $x$-momentum can be 
expressed as
\begin{eqnarray}
\rho u_j&=&f_1^j-f_3^j+f_5^j-f_6^j-f_7^j+f_8^j \mbox{\hspace{2in}}\nonumber \\
   &=&\frac{2\rho }3u_j+\frac{\rho}{6\tau }[u_{j-1}+u_{j+1}] 
          +\frac{\rho}{2\tau}[u_{j-1}v_{j-1}-u_{j+1}v_{j+1}] \nonumber \\
   & &   +(1-\frac 1{\tau}) 
          [f_5^{j-1}-f_6^{j-1}-f_7^{j+1}+f_8^{j+1}] \nonumber \\
   &=&\frac{2\rho }3u_j+\frac{\rho}{6\tau }[u_{j-1}+u_{j+1}]
          +\frac{\rho}{2\tau}[u_{j-1}v_{j-1}-u_{j+1}v_{j+1}] \nonumber \\
   & &+(1-\frac 1{\tau})
           [\rho u_{j-1}+\rho u_{j+1}-(f_1^{j-1}-f_3^{j-1}-f_7^{j-1}+f_8^{j-1})
      \nonumber \\
   & &              -(f_1^{j+1}-f_3^{j+1}+f_5^{j+1}-f_6^{j+1})]\nonumber \\
   &=&\frac{2\rho }3u_j+\frac{\rho}{6\tau }[u_{j-1}+u_{j+1}]
          +\frac{\rho}{2\tau}[u_{j-1}v_{j-1}-u_{j+1}v_{j+1}] \nonumber \\
    & &      +\frac 13 (1-\frac 1{\tau})
           [\rho u_{j-1}+\rho u_{j+1}-\rho u_j] .
\label{eq:rhouj}
\end{eqnarray}
which further gives us
\begin{equation}
\frac {u_{j+1}v_{j+1}-u_{j-1}v_{j-1}}{2 \delta}=\nu \frac{u_{j+1}+ u_{j-1}-2 
 u_j}{\delta^2} ,
\label{eq:diff}
\end{equation}
where $\nu=(2\tau -1) \delta/6 $ is the kinematic viscosity of the fluid
\cite{hou1}.
The above equation is exactly the second-order
finite-difference form of the simplified  incompressible
Navier-Stokes equation under the assumption Eq.~(\ref{eq:stead}) and constant
pressure:
\begin{equation}
 \frac{\partial (uv)}{\partial y}
=\nu \frac{\partial^2 u}{\partial y^2} 
\label{eq:ns}
\end{equation}

In th y-direction, it is easy to prove that
\begin{equation}
v_j=const.
\end{equation}
This result is obvious for an incompressible flow under the assumption 
Eq.(\ref{eq:stead}).

\section{Boundary Condition}
The derivation  of Eq.~(\ref{eq:rhouj}) is for the interior of the flow.
The same procedure can be used to derive the relationship of velocities near 
the wall.  For example, at $j=1$ (near the bottom of the flow region), we have
\begin{eqnarray}
\rho u_1&=&f_1^1-f_3^1+f_5^1-f_6^1-f_7^1+f_8^1 \nonumber  \\
   &=&\frac{2\rho }3u_1+\frac{\rho}{6\tau }[u_0+u_2]
          +\frac{\rho}{2\tau}[u_0v_0-u_2v_2]
      +(1-\frac 1{\tau})
          [f_5^0-f_6^0-f_7^2+f_8^2]  \nonumber \\
   &=&\frac{2\rho }3u_1+\frac{\rho}{6\tau }[u_0+u_2]
          +\frac{\rho}{2\tau}[u_0v_0-u_2v_2]  \nonumber \\
   & &+(1-\frac 1{\tau})
           [\rho \tilde{u}_0+\rho u_2-(f_1^0-f_3^0-f_7^0+f_8^0)
                           -(f_1^2-f_3^2+f_5^2-f_6^2)] \nonumber \\
   &=&\frac{2\rho }3u_1+\frac{\rho}{6\tau }[u_0+u_2]
          +\frac{\rho}{2\tau}[u_0v_0-u_2v_2]
      +\frac 13 (1-\frac 1{\tau})
           [\rho u_0+\rho u_2-\rho u_1]+ \nonumber \\
   & & (1-\frac 1{\tau})\rho [\tilde{u}_0 - u_0] ,
\label{eq:rhou1}
\end{eqnarray}
where $\tilde{u}_0 \equiv (f_1^0-f_3^0+f_5^0-f_6^0-f_7^0+f_8^0)/\rho$.
In the above derivation, it is assumed that equilibrium distributions 
at the boundary are calculated by using $u_0, v_0$. 
Eq.~(\ref{eq:rhou1}) further gives 
\begin{equation}
\frac {u_2v_2-u_0v_0}{2 \delta}=\nu \frac{u_2+ u_0-2 u_1}
{\delta^2} + \frac{\tau -1}{\delta}[\tilde{u}_0-u_0] .
\label{eq:diff2}
\end{equation}
Notice that $ \tilde{u}_0$
may not be equal to $u_0$ for a specific boundary condition.
Therefore, for
any $\tau \ne 1 $ and $ \tilde{u}_0 \neq u_0$, 
the velocity no longer satisfies the second-order difference equation
Eq.~(\ref{eq:diff})
at the fluid layer closest to the wall.  Nevertheless,
if we carefully choose a boundary condition to force $\tilde{u}_0=u_0$, 
the velocities will satisfy the same second-order difference equation as 
an approximation of the Navier-Stokes equation.

Similarly, near the top of the region, we have an equation: 
\begin{equation}
\frac {u_nv_n-u_{n-2}v_{n-2}}{2 \delta}=\nu \frac{u_n+ u_{n-2}-2 u_{n-1}}
{\delta^2} + \frac{\tau -1}{\delta}[\tilde{u}_n-u_n] ,
\label{eq:diff3}
\end{equation}
where $\tilde{u}_n 
\equiv  (f_1^n-f_3^n+f_5^n-f_6^n-f_7^n+f_8^n)/\rho$.
Again, the velocities will satisfy the same second-order difference equation 
if $\tilde{u}_n = u_n$.

The above restriction to $ \tilde{u}_0 $ (as well as $\tilde{u}_n$)
 has a profound physical meaning.  If
we analyze the momentum exchange between a wall and its nearest fluid layer,
we have (note that the streaming step is applied to the $f_i$ after
relaxation)
\begin{eqnarray}
\Delta M &=&f_5^1-f_6^1-(f_8^0-f_7^0)  \nonumber \\
         &=&(1-\frac 1{\tau})[f_5^0-f_6^0]+\frac {\rho}{6\tau}(u_0+3u_0v_0)-
         (f_8^0-f_7^0)  \nonumber \\
         &=&f_5^1-f_6^1-(1-\frac 1{\tau})[f_8^1-f_7^1]-\frac {\rho}{6\tau}
         (u_1-3u_1v_1) \nonumber .
\label{eq:delm}
\end{eqnarray}
Addition of the last two equations gives 
\begin{eqnarray}
2\Delta M &=&\Delta M +(1-\frac 1{\tau})[f_5^0-f_6^0-(f_8^1-f_7^1)]-\frac 
 {\rho}{6\tau}(u_1-u_0)+\frac {\rho}{2\tau}(u_1v_1+u_0v_0) \nonumber \\
          &=&(2-\frac 1{\tau})\Delta M -\frac 16(2-\frac 1{\tau})(u_1-u_0)
    +\frac {\rho}{2\tau}(u_1v_1+u_0v_0)+ \nonumber \\
          & & (1-\frac 1{\tau})\rho(\tilde{u}_0-u_0),
\label{eq:2delm}
\end{eqnarray}
or 
\begin{equation}
\Delta M = -\rho \nu \frac{u_1-u_0}{\delta} + 
\frac {\rho}2 (u_1v_1+u_0v_0)+(\tau-1) \rho(\tilde{u}_0-u_0) .
\label{eq:delmsol}
\end{equation}
where $(u_1-u_0)/\delta \approx \partial u/\partial y$ near the wall.
If $\tilde{u}_0 = u_0 $, this is the statement that the momentum exchange 
calculated from the exchange of  $f_i$'s is equal to the momentum exchange
carried by vertical velocity plus the viscous force. This is a 
momentum exchange principle.
Obviously, for any $\tau \ne 1$, if a specific boundary conditions gives 
$\tilde{u}_0 \ne u_0 $, it gives the wrong momentum exchange between a wall
and its nearest fluid neighbor.  In other words, in order to guarantee
the correct momentum exchange, one must choose a boundary condition
in which $\tilde{u}_0 = u_0 $ is satisfied.

The same conclusion can be drawn for boundary condition in the 
vertical direction.
Although we have shown in the last section that $v_j=const$
 in the interior of flow domain, this constant does not necessarily equal
to the vertical velocity at boundaries.
For instance, at $j=1$, we have
\begin{eqnarray}
\rho v_1&=&f_2^1-f_4^1+f_5^1+f_6^1-f_7^1-f_8^1 \nonumber  \\
        &=&\frac {\rho}{2\tau}[v_0+v_2]+\frac {\rho}{2\tau}[v_0^2-v_2^2]
           +(1-\frac 1{\tau})\rho[\tilde{v}_0+v_2-v_1] \nonumber
\end{eqnarray}
or
\begin{equation}
(2\tau -1)(v_0-v_1)+(v_0^2-v_1^2)+2(\tau-1)(\tilde{v}_0-v_0)=0,
\end{equation}
where $\tilde{v}_0 \equiv (f_2^0-f_4^0+f_5^0+f_6^0-f_7^0-f_8^0)/\rho$.
Obviously for any $\tau \neq 1$, the mass conservation is satisfied 
near the boundary ($v_1=v_0$) if and only 
if $\tilde{v}_0-v_0=0$.

Based on the analysis given above,
we propose a boundary condition for a straight boundary with given
velocity $u_0, v_0$ for the square lattice.
We will illustrate this at the bottom  boundary ($j=0$) as an example. 
After streaming, $f_0^0,f_1^0, 
f_3^0, f_4^0, f_7^0, f_8^0$ are known,  and 
$f_2^0, f_5^0, f_6^0$, hence $\rho$,  need to be defined. 

\begin{itemize}
\item Step 1: Calculation of the density $\rho$.

The restrictions on the boundary velocity and density discussed above
state that 
\begin{equation}
 \rho = f_0^0+f_1^0+f_2^0+f_3^0+f_4^0+f_5^0+f_6^0+f_7^0+f_8^0 , 
\label{eq:rho} 
\end{equation}
\begin{equation}
 \rho u_0  =  f_1^0-f_3^0 +f_5^0-f_6^0-f_7^0+f_8^0,  \mbox{\hspace{.9in}}
\label{eq:u0} 
\end{equation}
\begin{equation}
 \rho v_0  =  f_2^0-f_4^0+f_5^0 + f_6^0-f_7^0-f_8^0,\mbox{\hspace{.9in}}
\label{eq:v0}
\end{equation}

Comparison of Eqs.~(\ref{eq:rho}, \ref{eq:v0}) gives $\rho$ as:
\begin{equation}
 \rho =  \frac{1}{1-v_0} [f_0^0+f_1^0+f_3^0+2(f_4^0+f_7^0+f_8^0)].
\label{eq:rhod}
\end{equation}
\par
\item Step 2: Equilibrium part of the unknown distributions.

From the given boundary velocity and the density calculated in step 1,
we can calculate the equilibrium part of the distributions
$f_2^0, f_5^0, f_6^0$ by using Eq.~(\ref{eq:equil}).
\par
\item Step 3: Non-equilibrium part of the unknown distributions.

Substitution of $ f_i = f_i^{0 (eq)} + f_i^{0'}, i = 1,\cdots,8$ into
Eqs.~(\ref{eq:u0}, \ref{eq:v0}) gives
\begin{equation}
 f_5^{0'}-f_6^{0'} = -(f_1^{0'}-f_3^{0'}-f_7^{0'}+f_8^{0'}) ,
\end{equation}
\begin{equation}
 f_5^{0'}+f_6^{0'} = -(f_2^{0'}-f_4^{0'}-f_7^{0'}-f_8^{0'}) ,
\end{equation}
where $ f_i^{0'} $ is 
the non-equilibrium  distribution part of $f_i^0$.

Furthermore, we assume the bounce-back rule is still correct for the
non-equilibrium part of the particle distribution normal to
the boundary (in this case, $f_2^{0'}=f_4^{0'}$).
Under this condition, the other two undefined non-equilibrium distributions 
can be uniquely determined
\begin{equation}
 f_5^{0'} = f_7^{0'}-0.5(f_1^{0'}-f_3^{0'}) ,
\end{equation}
\begin{equation}
 f_6^{0'} = f_8^{0'}+0.5(f_1^{0'}-f_3^{0'}) .
\end{equation}
\end{itemize}
\par
The introduction of non-equilibrium part is only for the purpose of discussion
of the method, the non-equilibrium part need not be calculated explicitly.
In summary, the final form of the
unknown distributions from steps 2,3 can be determined as
\begin{eqnarray}
f_2^0&=&f_4^0+\frac {2 \rho v_0}3 ,\\
f_5^0&=&f_7^0-0.5(f_1^0-f_3^0)+\frac { \rho u_0}2 +\frac { \rho v_0}6, \\
f_6^0&=&f_8^0+0.5(f_1^0-f_3^0)-\frac { \rho u_0}2 +\frac { \rho v_0}6. 
\end{eqnarray}
Once all the after-streaming distributions at the boundary are determined,
the collision step can be easily applied to all nodes including the boundary
nodes.

The above procedures have specified a new boundary condition for any
given velocity straight boundary.
This boundary condition  produces the velocity  profile of the exact
solution for the plane Poiseuille flow with forcing \cite{he1}.
For the special flow with Eq.~(\ref{eq:stead}), this boundary condition
yields the correct velocity profile as the solution of the difference equation
Eqs.~(\ref{eq:diff}, \ref{eq:diff2}, \ref{eq:diff3})
under the condition $\tilde{u}_0 = u_0$, and $\tilde{u}_n = u_n$ (the formula
of the velocity profile is given in the following section).

\section{Velocity Profile}
The governing equation for the velocity profile Eq.~(\ref{eq:diff})
 can be solved under  different tangent velocities at upper and
bottom boundaries.
For simplicity, we assume $v_j = v_0 =const$, and $u_0$ and $u_n$ are given.
Eq.~(\ref{eq:diff}) can be written as
\begin{equation}
 (2 - R) u_{j+1} - 4 u_j + (2 + R) u_{j-1} = 0 ,
\label{eq:diff1}
\end{equation}
where $R \equiv v_0 \delta/\nu$.
Assuming a solution of the form
\begin{equation}
 u_j = \lambda^j ,
\label{eq:ujlam}
\end{equation}
we have a quadratic equation for $\lambda$:
\begin{equation}
 (2 -R)  \lambda^2
  - 4 \lambda + (2+R)  = 0 ,
\label{eq:lameq}
\end{equation}
 which has two solutions:
\begin{equation}
 \lambda_0 = 1, \ \ \lambda_1 = \frac{2+R }{2-R} . 
\label{eq:lambda}
\end{equation}
The general solution of Eq.~(\ref{eq:diff1}) is:
\begin{equation}
 u_j = a \lambda_1^j + b ,   \ \ j = 1, \cdots, n-1 ,
\label{eq:gen}
\end{equation}
where $a, b$ are some constants.
The corresponding difference equations near bottom and top boundaries,
Eqs.~(\ref{eq:diff2}, \ref{eq:diff3}),  are used to determine $a, b$.
These two equations  can be written in the
following way:
\begin{equation}
 (2-R) u_2 - 4 u_1 + (2+R) u_0 + \epsilon_0= 0 , \ \ 
\mbox{where}\ \ \epsilon_0 \equiv \frac{12(\tau-1)}{2\tau-1} (\tilde{u}_0 -u_0),
\label{eq:diffb}
\end{equation}
\begin{equation}
 (2-R) u_n - 4 u_{n-1} + (2+R) u_{n-2} + \epsilon_n= 0 , \ \ 
\mbox{where}\ \ \epsilon_n \equiv \frac{12(\tau-1)}{2\tau-1} (\tilde{u}_n -u_n),
\label{eq:difft}
\end{equation}
Substituting the general solution in Eq.~(\ref{eq:gen}) into 
Eqs.~(\ref{eq:diffb}, \ref{eq:difft}) yields a linear system of equations for
$a,b$, and solving them finally gives the solution:
\begin{equation}
 u_j= \frac{\lambda_1^j-1}{\lambda_1^n-1} u_n + 
  \frac{\lambda_1^n-\lambda_1^j}{\lambda_1^n-1} u_0 +
  \frac{\lambda_1^n-\lambda_1^j}{\lambda_1^n-1} \frac{\epsilon_0}{2+R}+
  \frac{\lambda_1^j-1}{\lambda_1^n-1} \frac{\epsilon_n}{2-R} .
\label{eq:uj}
\end{equation}
The last two term represents the error introduced at the boundaries 
if $\tilde{u}_0 \neq u_0$ or
 $\tilde{u}_n \neq u_n$.
 For the boundary condition introduced in
Section 3,  $\tilde{u}_0 = u_0, \tilde{u}_n = u_n$, hence
$\epsilon_0 = \epsilon_n = 0$.

In the special case  of 
 $\tilde{u}_0 = u_0 = 0$ and $\tilde{u}_n = u_n=U$, the solution becomes:
\begin{equation}
 \frac{u_j}{U} = \frac{\lambda_1^j-1}{\lambda_1^n-1} . 
\label{eq:ujU}
\end{equation}
It is easy to prove that this solution is a second-order approximation of 
the analytical solution 
\begin{equation}
\frac{u}{U}=\frac{e^{{\rm Re} \ y/L} -1}{e^{{\rm Re}}-1 } ,
\label{eq:solnu}
\end{equation}
where Re is the Reynolds number defined as Re$ = v_0 L/\nu$ with  $L$
being the width of the flow region.

\section{Discussion}

It is shown in the paper
that the flow velocity from the 2D LBGK  simulation in the injected porous
boundary case  satisfies  a second-order difference formula as an approximation
of the Navier-Stokes equation. This gives us a better understanding of 
the LBGK method.
The velocity formula near boundaries are consistent with the velocity 
formula inside the flow domain if the boundary condition is such that
the distribution functions at the boundary give the correct
boundary velocity and thereafter the momentum exchange principle is satisfied.
A boundary condition for any given velocity boundary for the 2D square lattice
LBGK model can be proposed based on the analysis in this paper.
This boundary condition can be easily extended to the 3D 15-velocity
direction model.

For the triangular lattice, the boundary condition
proposed by Nobel {\it et al.}
\cite{nobel} gives
the correct boundary velocity. It uses three equations:
$\sum_{i=0}^{6} f_i = \rho,$  and
$\sum_{i=1}^{6} f_i {\bf e}_i = \rho {\bf u}$ at a boundary node to
determine three unknowns:
$\rho$ and two $f_i$'s which are not defined after the streaming step
 (for example, $f_2, f_3$ at the bottom boundary).
The technique  does not have an immediate
extension to the square lattice because the number of unknowns on the
boundary in the
square lattice is larger than the number of restricting equations.
For the 3D 15-velocity direction lattice, Maier {\it et al.}
 \cite{bob} proposed
a boundary condition for  solid boundary (can be easily extended to the case
with boundaries with tangential velocities). First, the bounce-back is used
to find the unknown $f_i$'s, hence
the normal velocity is zero and the density and tangent velocity at the
wall node can be calculated, then the $f_i$'s, which are on the
directions pointing into the flow and have a non-zero projection on the wall,
will be adjusted to  correct the tangent velocity while keeping the
normal velocity and density unchanged. 
In the case where  the normal velocity is zero, the boundary condition 
proposed in this paper is
reduced to that proposed by Maier {\it et al.}. 
In the case where the normal velocity is non-zero,
the boundary condition in this paper can be extended to the 3D 15-velocity
direction lattice.
In the present discussion, analysis of the velocity profile is also given 
and 
the analysis provides a framework to study any boundary condition 
theoretically.

\section{Acknowledgments}

Discussions with R. Maier, R. Bernard, D. Grunau, L. Luo and H. Yang
are appreciated. 
Q.Z. would like to thank the Associated Western Universities  Inc.
for providing a fellowship and to thank G. Doolen and S. Chen 
for helping to arrange his visit to
 the Los Alamos National Lab.

\vfill\eject

\newpage
\section{Figure Caption}
Schematic plot of flow driven by the movement of upper boundary
with vertical injection fluid at the porous boundaries.
Also included is a 9-bit square lattice used in this paper.

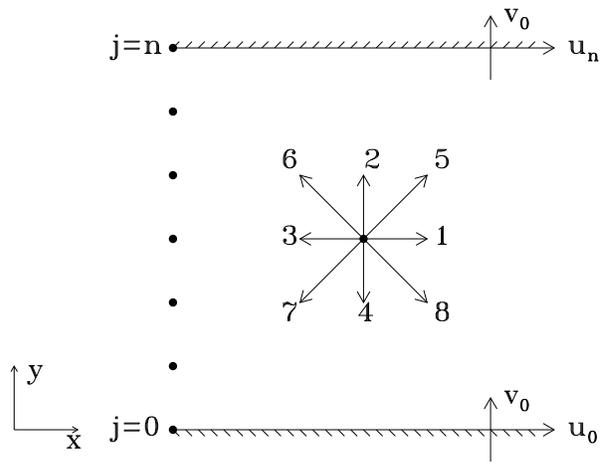
\end{document}